\documentclass[twocolumn,pra,aps,showpacs,superscriptaddress,floatfix,showkeys,nofootinbib,10pt]{revtex4-1}

\usepackage{amsmath,amsfonts,amssymb,amsthm}
\usepackage{color,graphicx}
\usepackage{bbm,bm}
\usepackage{booktabs}
\usepackage{multirow}
\usepackage{subcaption}

\usepackage{pifont}
\newcommand{\cmark}{\ding{51}}%
\newcommand{\xmark}{\ding{55}}%

\newtheorem{result}{Result}
\newtheorem{lemma}{Lemma}
\newtheorem{proposition}{Proposition}
\newtheorem{observation}{Observation}

\def\begeq{\begin{equation}}
\def\endeq{\end{equation}}
\def\begres{\begin{result}}
\def\endres{\end{result}}
\def\beglem{\begin{lemma}}
\def\endlem{\end{lemma}}
\def\begprop{\begin{proposition}}
\def\endprop{\end{proposition}}
\def\begobs{\begin{observation}}
\def\endobs{\end{observation}}

\newcommand{\la}{\langle}
\newcommand{\ra}{\rangle}

\newcommand{\tr}{\text{tr}}

\newcommand{\SP}{\text{ }}

\newcommand{\rw}{\rightarrow}
\newcommand{\RR}{\mathbbm{R}}
\newcommand{\CC}{\mathbbm{C}}
\newcommand{\SSS}{\mathbb{S}}
\newcommand{\ee}{\mathcal{E}}
\newcommand{\dd}{\mathcal{D}}
\newcommand{\ket}[1]{ | \, #1 \rangle}
\newcommand{\bra}[1]{ \langle #1 \, |}

\newcommand{\bk}[2]{\langle #1 \, | \, #2 \rangle}
\newcommand{\Ab}[1]{ \left| #1 \, \right|} 
\newcommand{\iu}{{i\mkern1mu}} 
\newcommand{\ba}{\begin{aligned}}
	\newcommand{\ea}{\end{aligned}}
\DeclareMathOperator{\Tr}{Tr}

\begin{document}
	
	\title{Connections Between Mutually Unbiased Bases and Quantum Random Access Codes}
	
	\author{Edgar A. Aguilar} \email{ed.alex.aguilar@gmail.com}\affiliation{Institute of Theoretical Physics and Astrophysics, National Quantum Information Center, Faculty of Mathematics, Physics and Informatics, 80-308, Gdansk, Poland}
	\author{Jakub J. Borka{\l}a} \affiliation{Institute of Theoretical Physics and Astrophysics, National Quantum Information Center, Faculty of Mathematics, Physics and Informatics, 80-308, Gdansk, Poland}
	\author{Piotr Mironowicz} \email{piotr.mironowicz@gmail.com} \affiliation{Department of Algorithms and System Modeling, Faculty of Electronics, Telecommunications and Informatics, Gda\'nsk University of Technology} \affiliation{National Quantum Information Centre in Gda\'nsk, 81-824 Sopot, Poland}
	\author{Marcin Paw{\l}owski} \affiliation{Institute of Theoretical Physics and Astrophysics, National Quantum Information Center, Faculty of Mathematics, Physics and Informatics, 80-308, Gdansk, Poland}

	\date{\today}
	
	\begin{abstract}
		We present a new quantum communication complexity protocol, the promise--Quantum Random Access Code, which allows us to introduce a new measure of unbiasedness for bases of Hilbert spaces. The proposed measure possesses a clear operational meaning and can be used to investigate whether a specific number of mutually unbiased bases exist in a given dimension by employing Semi--Definite Programming techniques. 
	\end{abstract}

	\maketitle
	
	\textit{Introduction.-}
	Mutually unbiased bases (MUBs) play a special role in the formalism of quantum mechanics. In particular they serve as complementary quantum tests, and find wide applicability in many fields of quantum information science such as quantum state tomography \cite{Ivo,WF}, quantum key distribution \cite{BB84}, quantum teleportation and dense coding \cite{BBELTZ}. Hence, a general understanding of MUBs is well motivated and of general interest, see \cite{DEBZ_review} for an extensive review and further references.
	
	Explicitly, two orthonormal bases $\{|\psi^1_i\ra\}_i$ and $\{|\psi^2_j\ra\}_j$ of $\CC^d$ are said to be mutually unbiased if 
	\begeq
		\label{eq:mub}
		\Ab{ \bk{\psi^1_i}{\psi^2_j} }^2 = \frac{1}{d} , \SP \forall i,j \in [d],
	\endeq
	where $[d] \equiv \{1,2,\dots,d\}$. The term unbiased is used because if we pick any basis vector $|\psi^1_i\ra$, then performing a measurement in the $\{|\psi^2_j\ra\}_j$ basis will yield a completely random result (i.e. each outcome $|\psi^2_j\ra$ will have equal detection probability $1/d$). 
	
	A set of MUBs in dimension $d$ is said to be \textit{maximal}, if there are $d+1$ bases which are all pairwise mutually unbiased. The construction of maximal sets when $d=p$, a prime number, was described by Ivonovic \cite{Ivo}, and later by Wootters and Fields when $d=p^k$, a prime power \cite{WF}. The general problem of whether $d+1$ bases exist for arbitrary dimensions remains open for at least the past 29 years. 
	
 	In particular it is an open question whether a complete set of MUBs exist even in the simplest case, namely in dimension 6. Zauner's conjecture states that \textit{no more than three MUBs exist in dimension 6} \cite{Zauner}. The task of proving the conjecture is a research field on its own, see e.g. \cite{JMMSW,Grassl} for partial analytical results supporting the conjecture. Numerical approaches have also failed to be conclusive, \cite{BW}.
		
	In this paper we introduce a novel protocol named promise-Quantum Random Access Code (pQRAC). The main idea of this protocol is to use the so-called $n^d \rightarrow 1$ Quantum Random Access Codes (QRACs) with certain constraints. Our main technical result shows that a specific average success probability of the protocol can be achieved if and only if $n$ MUBs exist in dimension $d$. 
	 	 
	The protocol allows us to create a new measure of unbiasedness, which quantifies the amount by which two (or more) bases are mutually unbiased. Other measures currently exist and are in use \cite{distance}, yet the presented one possesses a direct operational interpretation as the success probability of a well--defined communication task.
	
	 Furthermore, the pQRAC game is suitable for  numerical optimization techniques like Semi--Definite Programming (SDP)\cite{sdp}. In particular, one may use the see-saw method \cite{seesaw1} to search for $n$ MUBs in dimension $d$. What is more, pQRACs may be used together with the Navascues and Vertesi method \cite{PhysRevLett.115.020501} to discard the existence of $n$ MUBs in a particular dimension.  This exclusion is a rigorous statement, in contrast to drawing the conclusion out of the failure of trying to find them. As a proof of principle, we have applied our method to exclude the existence of 5 MUBs in dimension 3, and 6 MUBs in dimension 4. We have been unable to rule out the existence of 4 MUBs in dimension 6, but argue that the problem is now at arm's length for future researchers.

	\textit{Methods.-}
	We begin by introducing \textit{Random Access Codes} (RACs)\cite{ALMO_qracs}. An $n^d\rw1$ RAC is a protocol in which Alice tries to compress an $n$-dit string into $1$ dit, such that Bob can recover any of the $n$ dits with high probability. More precisely, Alice receives a uniformly distributed random input string $\mathbf{x}=x_1x_2\cdots x_{n}$, $x_i\in [d]$. She then uses an encoding function $\ee_c:[d]^n\rw [d]$ (possibly classically probabilistic), and is allowed to send one dit $a=\ee_c(\mathbf{x})$ to Bob. On the other side, Bob receives an input $y\in [n]$ (uniformly distributed), and together with Alice's message $a$ uses one of $n$ (possibly classically probabilistic) decoding functions $\dd_c^y:[d]\rightarrow [d]$, to output $b=\dd_c^y(a)$ as a guess for $x_y$. If Bob's guess is correct (i.e. $b=x_y$) then we say that they are \textit{successful}, otherwise we say that they are \textit{unsuccessful} or \textit{fail}.

	Similarly, we define $n^d\rw 1$ Quantum Random Access Codes (QRACs) where Alice encodes her input $n$-dit string into a $d$-dimensional quantum system (qudit) via $\ee_q:[d]^n\rw \CC^d$, and sends the qudit $\rho_{\mathbf{x}} = \ee_q(\mathbf{x})$ to Bob. He then performs one of his decoding functions $\dd_q^y:\CC^d \rw [d]$ to output his guess $b$ for $x_y$. The decoding function is simply a quantum measurement, i.e. he outputs his guess $b$ with probability $\mathbb{P}(b=x_y)=\tr[\rho_{\mathbf{x}} M^y_b]$, where the operators $M^y_b$ are POVMs (i.e. positive and $\forall y \SP \sum_b M^y_b = \openone$). As a figure of merit, we employ the optimal average success probability for both RACs and QRACs:
	\begeq
		\label{eq:qasp}
		\bar{P}_{c,q} (n,d) = \max_{\{\ee,\dd\}} \frac{1}{nd^n} \sum_\mathbf{x}\sum_y  \mathbb{P}(b=x_y) \SP .
	\endeq
	
	The maximization is over encoding-decoding strategies $\{\ee_{c,q},\dd_{c,q}\}$ (classical or quantum respectively), and the average is taken over all possible inputs $(\mathbf{x},y)$ of Alice and Bob.	In the quantum case, the optimal average success probability $\bar{P}_q$, can be achieved with pure states, $\rho_{\mathbf{x}} = |\mathbf{x}\ra\la \mathbf{x}|$ \cite{ALMO_qracs}, where $|\mathbf{x}\ra$ is the eigenvector of $\sum_y M^y_{x_y}$ with largest eigenvalue. In \cite{Farkas}, it was shown that for $2^d\rw 1$ QRACs this maximum is achieved when the operators $M^y_b$ are (rank 1) projective measurements. Therefore, throughout the rest of this letter we will be considering only pure-state encoding and von-Neumann measurements.
	
	RACs and QRACs have increasingly become an experimental tool to test the ``quantumness" or non-classical behavior of a system \cite{THMB_expqrac,QRACexp}. For fixed $n$ and $d$, we have $\bar{P}_c < \bar{P}_q$, and a gap is exploited to show that a system is behaving non-classically . For example a $2^2\rw1$ RAC has $\bar{P}_c=0.75$, while the corresponding QRAC has an optimal average success probability of $\bar{P}_q=(2+\sqrt{2})/4\approx 0.8536$ \cite{ANTV}. Thus for a system of dimension 2, observing an average success probability greater than $0.75$ indicates non-classical behaviour. 
	
	The quantum advantage comes from encoding Alice's state as a superposition of the bases $\{|\psi^1_i\ra\}_i$ and $\{|\psi^2_j\ra\}_j$, namely $|\mathbf{x}\ra = \alpha |\psi^1_{x_1}\ra + \beta |\psi^2_{x_2}\ra$, while Bob measures in the $\{|\psi^y_i\ra\}_i $ basis. We have the following:		
	\beglem 
		\label{result:1}
		For a $2^d\rw 1$ QRAC, the optimal average success probability 
		\begeq
			\label{eq:21optimal}
			\bar{P}_q (2,d) = \frac{1}{2}\left(1+\frac{1}{\sqrt{d}}\right) 
		\endeq
		 is obtained if and only if Bob's measurement bases $\{|\psi^1_i\ra\}_i , \{|\psi^2_j\ra\}_j$ are mutually unbiased.  
	\endlem
	The proof is given in Supplementary Material~\ref{sec:2d1}. We find it interesting to note here an observation that Lemma~\ref{result:1} cannot be generalized to the case of $n^d\rw1$ QRACs for $n \geq 3$, as stated below:
	\begobs
		The MU condition on Bob's measurement bases is not sufficient for obtaining the optimal average success probability in $n^d\rw1$ QRACs when $n \geq 3$.
	\endobs
	
	The proof of this result is by direct calculation (See Supplementary Material~\ref{sec:anomalies} for details). This occurs since there are inequivalent subsets of MUBs  (i.e. not related by unitary transformations) in higher dimensions. As an example, let us consider the case $n=3, d=5$. Bob must choose 3 different measurement bases, and he can do so in ${{6}\choose{3}} =20$ ways. Half of those selections lead to an average success probability of 0.610855, while the other half give 0.596449. Hence, the choice of the subset of MUBs matters. This feature occurs also for other choices of $n$ and $d$. However, we conjecture that the optimal average success probability for $n^d\rw1$ QRACs is indeed achieved with a suitable choice of MUBs. 
	
\begin{figure}
		\includegraphics[scale=0.7]{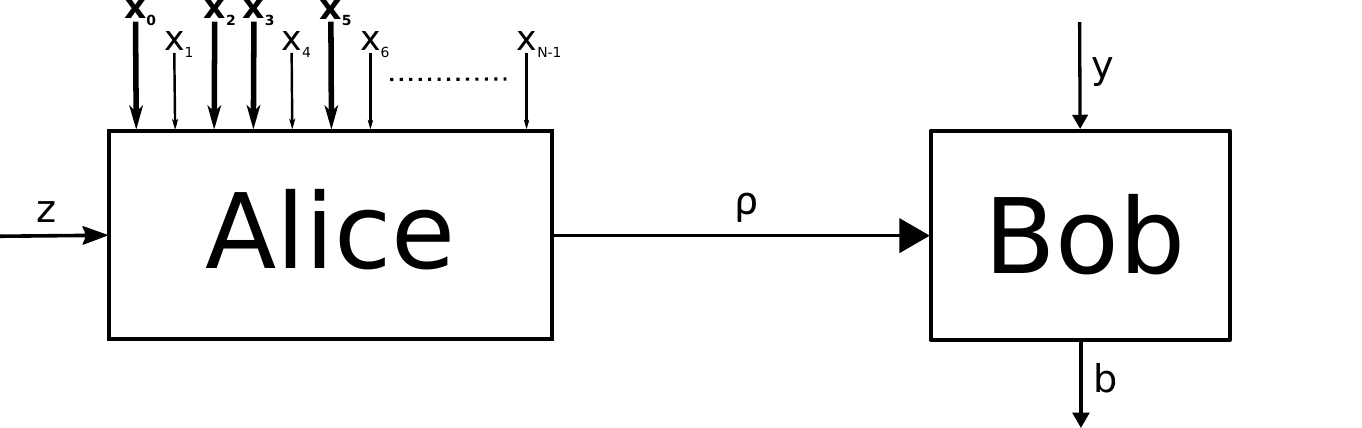}
		\caption{Schematic representation of a $(n,m)^d\rw1$ promise--Quantum Random Access Code. Here $x_i\in [d],y\in[n]$, and $z$ is a subset of $[n]$ with $m$ elements. The bold inputs $x_k$ depict $k\in z$. $\rho$ is the quantum state that Alice sends to Bob.}
		\label{fig:pQRACplot}
	\end{figure}

	Next we define a $(n,m)^d\rw1$ promise-QRAC (pQRAC), $m\leq n$, as an $n^d\rw1$ QRAC with an extra promise. Let $S^n_m$ be the set of all possible subsets of $[n]$ of size $m$. Then in a pQRAC, Alice receives an additional input $z \in S^n_m$, with the promise that $y \in z$. That is, Alice knows that Bob will not be questioned over some of Alice's inputs, see Fig.~\ref{fig:pQRACplot} for an illustration of a pQRAC. 
	
	Hence, the optimal average success probability \eqref{eq:qasp}, is modified in the case of $(n,m)^d\rw1$ pQRACs to:
	\begeq
		\label{pQRACasp}
		\tilde{P}_q (n,m,d) =\max_{\{\rho,\{M\}\}} \frac{1}{\binom{n}{m} md^m} \sum_{z\in S^n_m} \sum_{\mathbf{x}_z} \sum_{y\in z} \tr[\rho_{\mathbf{x},z} M^y_{x_y}],
	\endeq
	where the summation over $\mathbf{x}_z$ indicates a summation over $x_{i_1}, x_{i_2} , \dots, x_{i_m}$ such that $\{i_1, i_2, \dots , i_m\}= z$ , and the maximization is taken over all quantum encoding and decoding strategies $\{\rho,\{M\}\}$. Now, we are able to prove our main technical result:
	\beglem
		\label{result:2}
		For a $(n,2)^d\rw1$ pQRAC, the following holds:
		\begeq
			\tilde{P}_q(n,2,d) \leq \frac{1}{2}\left(1+\frac{1}{\sqrt{d}}\right) 
		\endeq
		with equality iff at least $n$ MUBs exist in dimension $d$.
	\endlem
	\begin{proof}
		We begin by writing the optimal average success probability of the  $(n,2)^d\rw1$ pQRAC.
		\begin{align*}
			\tilde{P}_q (n,2,d) &=\max_{\{\rho,\{M\}\}} \frac{1}{\binom{n}{2} 2d^2} \sum_{z\in S^n_2} \sum_{\mathbf{x}_z} \sum_{y\in z} \tr[\rho_{\mathbf{x},z} M^y_{x_y}] \\
			&\leq  \frac{1}{\binom{n}{2}} \sum_{z\in S^n_2} \left( \max_{\{\rho,\{M\}\}} \frac{1}{2d^2} \sum_{\mathbf{x}_z} \sum_{y\in z} \tr[\rho_{\mathbf{x},z} M^y_{x_y}] \right).
		\end{align*}
		The inequality follows, since the strategies to maximize the summands might not be compatible with each other globally. In fact, we recognize the term in parenthesis as $\bar{P}_q (2,d)$, the optimal success probability for a $2^d \rw 1$ QRAC \eqref{eq:qasp}. From Lemma \ref{result:1}, this maximization occurs if and only if the measurement bases corresponding to the set $z$ are mutually unbiased. It follows that it is possible to simultaneously satisfy all of these maximization constraints iff there exists $n$ MUBs in dimension $d$.		
	\end{proof}

	The intuition behind Lemma \ref{result:2}, is that Bob must be ready to measure in all $n$ bases. If there exist $n$ bases which are all pairwise mutually unbiased, then essentially they are just playing a more complicated version of the usual $2^d \rw 1$ QRAC. If these bases do not exist, then for some $z \in S^n_2$ , the protocol will not be able to achieve the optimal value \eqref{eq:21optimal}, dropping the entire average.

	\textit{Results.-}
	In the context of MUBs, reference \cite{BBELTZ} has introduced a distance measure between two bases $\{|\psi^1_i\ra\}_i$ and $\{|\psi^2_j\ra\}_j$  which quantifies unbiasedness:
	\begeq
		D^2_{\psi^1 \psi^2} = 1 - \frac{1}{d-1} \sum_{i,j\in [d]} \left(|\la \psi^1_i |\psi^2_j\ra|^2 -\frac{1}{d}\right)^2.
	\endeq
	The measure is symmetric ($D^2_{\psi^1 \psi^2} = D^2_{\psi^2 \psi^1}$). If the bases are the same, then $D^2_{\psi^1 \psi^1} = 0$. The maximum $D^2_{\psi^1 \psi^2} = 1$ is obtained iff the bases are mutually unbiased. For a set of $n$ bases in $\CC^d$ ($\{\psi^j\} = \{|\psi^j_i\ra\}_i$,  $j\in[n]$), one can analyze the average square distance between all possible pairs of bases \cite{distance}:
	\begeq
		\label{eq:dsq}
		\bar{D}^2(\{\psi^i\}_i)  = \frac{1}{\binom{n}{2}} \sum_{\{a,b\}\in S^n_2} D^2_{\psi^a \psi^b}.
	\endeq
	Likewise, $\bar{D}^2=1$ iff all bases are pairwise mutually unbiased. However, \eqref{eq:dsq} is an abstract distance measure, lacking an operational interpretation. 
	
	Lemma~\ref{result:2} immediately leads us to our first result. Given a set of $n$ bases in dimension $d$ we define as their unbiasedness measure the average success probability in a $(n,2)^d \rw 1$ pQRAC if the bases are used as Bob's measurement bases. This measure is thus defined operationally and has the following properties: (1) The maximum of $\bar{P}_q(2,d) = \frac{1}{2}(1+d^{-1/2})$ is attainable iff all bases are pairwise unbiased, (2) It is symmetric under permutation of bases, and (3) The minimal value of $\bar{P}_c(2,d) = \frac{1}{2}(1+d^{-1})$ is achieved iff all bases are the same. The optimal classical success probability of $n^d\rw 1$ RACs is shown in \cite{Czechlewski18}.

	Explicitly, given $n$ bases of $\CC^d$, $\{ \psi^i \}_i$, the maximum attainable average success probability of the pQRAC, $\bar{p}$, is:
	\begeq
		\label{eq:Pmeasure}
		\bar{p}(\{\psi^i\}_i) =  \frac{1}{\binom{n}{2}} \sum_{\{a,b\}\in S^n_2} \left( \frac{1}{2} + \frac{1}{2d^2} \sum_{i,j\in[d]}  |\la \psi^a_i | \psi^b_j \rangle | \right),
	\endeq
	which comes as a direct conclusion of Lemma~\ref{result:1}. We may normalize \eqref{eq:Pmeasure} such that the minimum value is $0$ (obtained iff all bases are the same), and the maximum value is $1$ (obtained iff all bases are pairwise MU) and get the expression:
	\begeq
		\label{eq:Qmeasure}
		\bar{Q}(\{\psi^i\}_i) = \frac{\bar{p}(\{\psi^i\}_i) - \bar{P}_c(2,d)}{\bar{P}_q(2,d) - \bar{P}_c(2,d)} .
	\endeq
See Supplementary Material \ref{sec:Qbar} to see a direct comparison between \eqref{eq:Qmeasure} and \eqref{eq:dsq}.

For illustrative purposes, we have optimized the value of the $(4,2)^6 \rw 1$ pQRAC game expression using the see-saw method \cite{seesaw1}. This allows us to show how the optimization of \eqref{eq:Qmeasure} may be used to construct MUBs in a particular dimension, as well as providing numerical examples of how $\bar{D}^2$ and $\bar{Q}$ compare. With this method, the maximal value of $\bar{D}^2$ of four MUBs in dimension $6$ we obtained is $0.998284$ with $\bar{Q}=0.998045$. On the other hand, the bases from \cite{BH, RLE} have $\bar{D}^2 = 0.998292$, and $\bar{Q} = 0.998036$. With this result one sees that the two measures, \eqref{eq:dsq} and \eqref{eq:Qmeasure} are not equivalent, and induce different partial orderings on the sets of bases. See Supplememntary Material \ref{sec:seesaw} for more details.
	
	Our second result is another direct application of Lemma \ref{result:2}, and deals with ruling out if there are $n$ mutually unbiased bases in dimension $d$. Explicitly, if it is possible to show that there are no sets of encoded states and measurement bases that would obtain a success probability of $\bar{P}_q(2,d)$, then one immediately concludes that there does not exist $n$ MUBs in the given dimension. Thus one may use the SDP hierarchy of relaxations proposed by Navascues and Vertesi (NV) \cite{PhysRevLett.115.020501}. The method defines a sequence of SDP problems yielding upper bounds to optimization tasks over quantum probability distributions with dimensional constraints. One can show that the method converges to the accurate quantum values \cite{PhysRevA.92.042117}. If at a given level of the hierarchy the upper bound falls below the threshold $\bar{Q}=1$, then the conclusion follows. We emphasize that if $n$ MUBs do not exist in a particular dimension, then applying the SDP hierarchy to the $(n,2)^d$ pQRAC gives an algorithmic way of proving their non-existence. On the other hand, if $n$ MUBs do exist, the proposed method will fail to draw a conclusion. 
	
	\textit{Implementing the hierarchy.-}
Let us try to directly apply the NV hierarchy to the $(n,2)^d \rw 1$ pQRAC. To implement the $k$-th level of the hierarchy, $Q^k$, the set $\SSS^k_d$ of all feasible moment matrices of order $2k$ arising from quantum systems of dimension $d$ must be calculated. For this, moment matrices $\Gamma^j_k$ are randomly generated from this set until $\text{span}(\{\Gamma^j_k\}_j) = \SSS^k_d$. In practice, the algorithm keeps creating new moment matrices $j=\{1,2,\ldots,v_k\}$ and stops when $\Gamma^{v_k+1}_k \in \text{span}(\{\Gamma^j_k\}_{j=1}^{v_k})$. The method requires an assumption on the rank of the projectors $\{M^y_b\}$, but in our scenario Bob's optimal strategy is to implement $d$-dimensional von Neumann measurements, therefore all operators are rank 1.

In order to generate $\Gamma^j_k$, we randomly choose $A=\binom{n}{2} d^2$ states for Alice to encode and $B=nd$ measurement operators for Bob ($n$ bases of $\CC^d$). Then, $\Gamma^j_k$ contains the traces of all strings of size less than or equal to $2k$ constructed from Alice's states and Bob's operators. For example, typical matrix elements of $\Gamma^j_1$ include $\tr[\rho^j_{\mathbf{x},z}\rho^j_{\mathbf{x}',z'}]$, $\tr[\rho^j_{\mathbf{x},z}M^{y,j}_b]$, and $\tr[M^{y,j}_b M^{y',j}_{b'} ]$. While in $\Gamma^j_3$, we can find $\tr[\rho^j_{\mathbf{x},z} M^{y,j}_{b} M^{y',j}_{b'} M^{y'',j}_{b''} \rho^j_{\mathbf{x}',z'} M^{y''',j}_{b'''}]$, etc. 

We write the $k$-th order relaxation to our problem as the following semidefinite program \cite{PhysRevLett.115.020501}: 
\begin{equation}
\begin{split}
\label{eq:SDPklevel}
&\tilde{P}_q(n,2,d) = \max \tr[ \hat{\mathcal{B}} \text{ }\Gamma_k] \\
&\text{s.t. } \Gamma_k \in \SSS^k_d , \text{ }\left(\Gamma_k\right)_{1,1} = 1, \text{ }\Gamma_k\geq 0,
\end{split}
\end{equation}
where we call $\hat{\mathcal{B}}$ the \textit{pQRAC game matrix}, and construct it to ``pick out" the values $\tr[\rho_{\mathbf{x},z} M^y_b]$ from $\Gamma_k$ such that $b=x_y$ and $y\in z$. 

Roughly $\frac{1}{2}(A+B)^{4k}$ real-valued numbers need to be stored in a computer's RAM in order to describe the set of all feasible moment matrices $\SSS^k_d$. Below, we describe a potentially quadratic reduction in the problem's memory requirements. See Supplementary Material \ref{sec:estimate} for details. 

Note that $\hat{\mathcal{B}}=\hat{\mathcal{B}}^T$, and is a sparse matrix with a lot of symmetries. In this case, we employ the symmetries corresponding to relabeling measurement device outputs, and the ones corresponding to permuting the labels of the measurement devices themselves. This approach has been followed on the NPA hierarchy in the Bell-test  scenario \cite{BellSymmetries}.


Let $\hat{\mathcal{B}}$ be invariant under the group of transformations $\mathcal{G}$. In other words, for every representation $G$ of an element $g\in\mathcal{G}$, $G \hat{\mathcal{B}} G^T = \hat{\mathcal{B}}$. Then, if we apply a group action on the game matrix inside the objective function \eqref{eq:SDPklevel}, this would be equivalent to applying a group action on $\Gamma_k$. Namely, $\tr[  \hat{\mathcal{B}}  \SP \Gamma_k] = \tr[ \hat{\mathcal{B}} \SP G^T \Gamma_k G]$. Therefore, it is unnecessary to consider the full space of feasible moment matrices $\SSS^k_d$ and can simplify \eqref{eq:SDPklevel} into:

\begin{equation}
\begin{split}
\label{eq:SDPklevelsymm}
&\tilde{P}_q(n,2,d) = \max \tr[ \hat{\mathcal{B}} \text{ }\hat{\Gamma}_k] \\
&\text{s.t. } \hat{\Gamma}_k \in \mathcal{G}(\SSS^k_d) , \text{ }\left(\hat{\Gamma}_k\right)_{1,1} = 1, \text{ }\hat{\Gamma}_k\geq 0 ,
\end{split}
\end{equation}
where we denote $\mathcal{G}(\SSS^k_d)$ as the set of feasible moment matrices which are $\mathcal{G}$-invariant. In order to implement this, we generate random invariant moment matrices $\hat{\Gamma}^j_k$ by first creating a moment matrix $\Gamma^j_k$ and averaging it out over all of the group elements:
\begeq
\label{eq:avgsymm}
\hat{\Gamma}^j_k = \frac{1}{|\mathcal{G}|} \sum_{\mathcal{G}} G\SP\Gamma^j_k G^T.
\endeq
Clearly $\hat{\Gamma}^j_k \in \mathcal{G}(\SSS^k_d)$, and this is repeated until $\text{span}(\{\hat{\Gamma}^j_k\}_j) = \mathcal{G}(\SSS^k_d)$. To illustrate the power of the proposed method, we report that for a $(4,2)^5\rw 1$ pQRAC, $\text{dim}(\SSS^1_5)=13672$ and the SDP running time was 22.5h on a desktop computer, whereas $\text{dim}(\mathcal{G}(\SSS^1_5))=7$ and had a total run-time of 50s. 	 
	
Using this, we have implemented $Q^1$ and a subset of the ``almost quantum" level \cite{Nav15q1ab} ($Q^{1+\textit{succ}}$) for some relevant pQRAC cases, see Table \ref{tab:tabSDP}. The level $Q^{1+\textit{succ}}$ includes traces of strings of length $\leq 2$ from the set of operators $\{ \{\rho^j_{\mathbf{x},z}\}, \{M^{y,j}_b\}, \{ \rho^j_{x_{z_1},x_{z_2},\{z_1,z_2\}} M^{z_i,j}_{x_{z_i}}\}\}$. That is, we also included pairs of states and measurements which lead to successful trials.  The details of the implementation are found in Supplementary Material \ref{sec:Symmetries}. As a proof of principle, we note that we have been able to rigorously rule out the existence of $d+2$ MUBs in dimensions $d=3$ and $d=4$. However, with our numerical precision and at this hierarchy level we have been unable to rigorously exclude the existence of 4 MUBs in dimension 6. 

We notice that the hierarchy level $Q^{1+\textit{succ}}$ was also unable to rule out the existence of 4 MUBs in $d=2$. If these four bases existed, together with the $d=3$ MUBs one could create four MUBs in dimension 6. We conjecture that in order for a level of the hierarchy to be able to rule out the existence of 4 bases in dimension 6, it must first rule out the existence of 4 MUBs in $d=2$. In future work, we wish to find more efficient ways of calculating \eqref{eq:SDPklevelsymm}, and higher levels of the hierarchy.
	
\begin{table}[t]				
		\begin{tabular}{|c|c|c|c|}
			\hline 
			case $(n,2)^d$ & $\bar{Q}$ for $Q^1$ & $\bar{Q}$ for $Q^{1+\textit{succ}}$ 
 \\  \hline  
			$(3,2)^2$ & $\textbf{0.999999}99$ & $\textbf{0.999999}99$ 
 \\  \hline  
			$(4,2)^2$ & $\textbf{0.999999}99$ & $\textbf{0.999999}92$ 
 \\  \hline  
			$(4,2)^3$ & $\textbf{1.131652}47$ & $\textbf{0.999999}95$ 
 \\  \hline  
			$(5,2)^3$ * & $\textbf{1.131652}49$ & $\textbf{0.999898}95$
 \\  \hline   
			$(5,2)^4$ * & $\textbf{1.242640}66$ & $\textbf{0.999999}96$
 \\  \hline			 
			$(6,2)^4$ * & $\textbf{1.242640}66$ & $\textbf{0.999916}55$
 \\  \hline 
            $(3,2)^6$ & $\textbf{1.428825}41$ & $\textbf{0.999999}98$ 
 \\  \hline 
		    $(4,2)^6$ * & $\textbf{1.428825}38$ & $\textbf{0.999999}97$ 
 \\  \hline 

		\end{tabular}
		\caption{Results of implementing hierarchy levels $Q^1$ and $Q^{1+\textit{succ}}$ to $(n,2)^d\rw 1$ pQRACs. The * indicates an optimization over $\mathcal{G}(\SSS^k_d)$. Other cases were executed for both $\SSS^k_d$ and $\mathcal{G}(\SSS^k_d)$ to test our code implementation.
		 \label{tab:tabSDP}}
	\end{table}

	\textit{Conclusions.-}
	In this paper we give a new class of quantum games, pQRACs, which serve as an operational way of testing unbiasedness. It also enables one to reformulate the problem of searching for a given number of MUBs in a particular dimension as a problem of optimizing the strategy of the pQRAC game. In particular if one is able to get a proper upper bound on the value of the game, then our formulation allows to exclude the existence of a given number of MUBs in the considered dimension. We have exploited the symmetries of the pQRAC game matrix in the Navascues and Vertesi hierarchy. We hope this will lead to rigorously proving Zauner's conjecture by considering higher levels.

	\textit{Acknowledgments.-}
	The paper was supported by EU grant RAQUEL, ERC AdG QOLAPS, National Science Centre (NCN) Grant No. 2014/14/E/ST2/00020 and DS Programs of the Faculty of Electronics, Telecommunications and Informatics, Gda\'nsk University of Technology. EA acknowledges support from CONACyT.
	The SDP optimizations have been performed using OCTAVE \cite{octave} with SDPT3 solver \cite{SDPT3a,SDPT3b}, SeDuMi \cite{Sturm98usingsedumi}, and YALMIP toolbox \cite{yalmip}. We thank D. Saha and M. Farkas for discussions, and I. Bengtsson for guidance with the literature.

	\bibliographystyle{ieeetr}
	\bibliography{QRACMUBrefs}

	\begin{appendix}
			
		\section{Optimal average success probability of $2^d\rightarrow 1$ QRAC}
		\label{sec:2d1}
		
		Let $\{|\psi_i\ra\}_i$ and $\{|\phi_j\ra\}_j$ be two orthonormal bases used on Bob's side to perform the von Neumann measurement, and we write their projectors as $|\psi_i\ra\la\psi_i|=M^0_i$ and $|\phi_j\ra\la\phi_j|=M^1_j$. The optimization of the average success probability is simply a maximization over the measurement bases $\{M^0_i\},\{M^1_j\}$ of the expression:
		\begeq
			\bar{p}_{q} =  \frac{1}{2d^2} \sum_{x_0,x_1=0}^{d-1} \tr[\rho_{x_0,x_1} (M^0_{x_0}+M^1_{x_1})].
		\endeq
		The maximum is achieved when $\rho_{x_0,x_1}$ (a pure state) is an eigenvector, corresponding to the largest eigenvalue of $(M^0_{x_0}+M^1_{x_1})$. This can be seen by writing $\rho_{x_0,x_1} = |\xi\ra\la\xi|$  , and expressing the pure state as $|\xi\ra = \sum c_k |k\ra$, where $\{|k\ra\}_k$ is the eigenbasis of the sum of the operators ($M^0_{x_0}+M^1_{x_1} = \sum \lambda_k |k\ra\la k|$). Then $\tr[\rho_{x_0,x_1} (M^0_{x_0}+M^1_{x_1})] = \sum |c_k|^2 \lambda_k $, which is clearly maximal when $\rho_{x_0,x_1} = |\lambda_{\text{max}}\ra\la \lambda_{\text{max}}|$.
		
		Thus we are trying to maximize the sum $\sum_X \lambda_{\max}(M^0_{x_0}+M^1_{x_1})$. Direct calculations considering only two dimensional subspaces at a time yields:
		\begeq
			\label{eq:21qracmax}
			\bar{P}_{q} =  \frac{1}{2d^2} \sum_{x_0,x_1=0}^{d-1} \left(1+\alpha_{x_0,x_1} \right),
		\endeq
		where $\alpha_{x_0,x_1}=\sqrt{\tr[M^0_{x_0}M^1_{x_1}]}=|\la\psi_{x_0}|\phi_{x_1}\ra|$ is simply the modulus of the inner product between two elements of the different bases.
			
		To see this, we first fix $x_0$ and $x_1$. Then, any 2 linearly independent vectors $\ket{\psi_{x_0}},\ket{\phi_{x_1}}$ span a 2 dimensional subspace of $\mathbb{C}^d$. Let $\ket{0},\ket{1}$ be the basis vectors of such a subspace and for simplicity we choose them such that:
		\begin{align}
			&\ket{\psi_{x_0}}=\ket{0} \\
			&\ket{\phi_{x_1}}=\alpha\ket{0} +\beta\ket{1},
		\end{align}
		where $|\alpha|^2+|\beta|^2=1$, and $\alpha=\alpha_{x_0,x_1}=|\la\psi_{x_0}|\phi_{x_1}\ra|$. Then, we solve the characteristic equation for the measurement operator, $0=\det(\ket{\psi_{x_0}}\bra{\psi_{x_0}}+\ket{\phi_{x_1}}\bra{\phi_{x_1}}-\lambda\mathbb{I})$, to get the following two solutions:
		\begin{equation}
			\lambda_{\pm}=1 \pm  \sqrt{1-|\beta|^2}=1\pm\alpha, 
		\end{equation}
		where $\lambda_+$ is the maximum eigenvalue used in \eqref{eq:21qracmax}.
				
		Before being able to maximize \eqref{eq:21qracmax}, we briefly introduce the concepts of majorization and of Schur concavity. We say that a probability distribution $\mathbf{p}=(p_1, p_2 , \ldots , p_N)$ \textit{majorizes} the probability distribution $\mathbf{q}=(q_1, q_2, \ldots, q_N)$, denoted by $\mathbf{p} \succ \mathbf{q}$, if:
		\begeq
		\sum_{i=1}^k  p_i^{\downarrow} \geq \sum_{i=1}^k q_i^{\downarrow} \text{ , } \forall k\in [N] , 
		\endeq 
where $\mathbf{p}^\downarrow $ $(\mathbf{q}^\downarrow)$ is a vector with the same components as $\mathbf{p}$ $(\mathbf{q})$, but written in descending order. A function $F:\RR^N \rw \RR$ is said to be Schur concave if $\mathbf{p} \succ \mathbf{q}$ implies $F(\mathbf{q}) \geq F(\mathbf{p})$. In particular, if $f:\RR\rw \RR$ is a concave function , then $F = \sum_i f(x_i)$ is Schur concave.

	Finally, note that $\sum_x \alpha_x^2=d$, is a constant. Let us define the probability distribution $\mathbf{a} = (a_1, a_2 ,\ldots , a_{d^2})$, with $a_i = \frac{1}{d} \alpha^2_i , \forall i$. Then \eqref{eq:21qracmax} may be rewritten as:
	\begeq
	\bar{P}_q = \frac{1}{2d^2} \sum_{x\in [d^2]} 1 + \sqrt{d a_x} .
	\endeq
By the remark above this function is Schur concave, and is therefore maximized by the uniform distribution $\mathbf{u} = \frac{1}{d^2} (1 , 1 , \ldots , 1)$ \cite{inequalities}. This is because $\mathbf{u}$ is majorized by all probability distributions.	Explicitly, we obtain $\alpha_x^2 = d^{-1} , ~ \forall x$, which is precisely the condition of unbiasedness \eqref{eq:mub}.

		\section{Insufficiency of MUBs for optimality in $3^d\rw 1$ QRACs}
		\label{sec:anomalies}
		
		In this section we would like to find the maximum success probability for $3^d\rw1$ QRACs, and prove that the maximum is achieved when using MUBs. To do this we need to find the optimal encoding/decoding strategy for Alice and Bob.\\
		
		For a given encoding/decoding strategy, the average success probability is given by \eqref{eq:qasp}, which we write out here as:
		\begeq
			\bar{p}=\dfrac{1}{3d^{3}}\sum_{i=0}^{d-1} \sum_{j=0}^{d-1} \sum_{k=0}^{d-1}  \tr [\rho_{ijk} (\Psi_{i} + \Phi_{j}+ \Theta_{k})],
		\endeq
		where $\Psi_i = |\psi_i\ra\la\psi_i|$, $\Phi_j=|\phi_j\ra\la\phi_j|$, $\Theta_k=|\theta_k\ra\la\theta_k|$ are three projectors which are used by Bob to measure the message from Alice (encoded as $\rho_{ijk}$).  To maximize the above expression, we will let Alice's encoding $\rho_{ijk}$ be a pure state corresponding to the eigenvector of the largest eigenvalue of the measurement operators $\lambda_{\text{max}}(\Psi_{i} + \Phi_{j}+ \Theta_{k})$. For the moment we analyze the eigenvalues of such operators, and drop the $i,j,k$ subscripts. In short, we want to solve the following characteristic equation:
		\begin{equation}
			\det(\ket{\psi}\bra{\psi}+\ket{\phi}\bra{\phi}+\ket{\theta}\bra{\theta}-\lambda\mathbb{I})=0,
		\end{equation}
		which leads to the following third degree polynomial for the eigenvalues: 
		
		\begin{subequations}
			\begin{align}
			&-\lambda^3+3\lambda^2+m\lambda+n=0, \\
			&m=\lvert\bra{\psi}\phi\ra\rvert^2 + \lvert\bra{\psi}\theta\ra\rvert^2 + \lvert\bra{\phi}\theta\ra\rvert^2-3,\\
			&n = \text{det}(\ket{\psi}\bra{\psi}+\ket{\phi}\bra{\phi}+\ket{\theta}\bra{\theta}).
			\end{align}
			\label{eq:char}
		\end{subequations}
		
		The 3 vectors $\ket{\psi},\ket{\phi},\ket{\theta}$ span at most a 3 dimensional subspace of $\mathbb{C}^d$. Let $\ket{0},\ket{1},\ket{2}$ be the basis vectors of such a subspace and  write:
		\begin{subequations}
			\begin{align}
			&\ket{\psi}=\ket{0}, \\
			&\ket{\phi}=\alpha\ket{0} +\beta\ket{1},\\
			&\ket{\theta}=a\ket{0} + b\ket{1}+ c\ket{2}.
			\end{align}
		\end{subequations}
		Then $n=|c\beta|^2$. We observe that the parameter $m\in [-3,0 ]$ and $n\in [0,1]$ but since $\ket{\psi}\bra{\psi}+\ket{\phi}\bra{\phi}+\ket{\theta}\bra{\theta}$ is a Hermitian operator, we need to focus only on the subdomain of $m$ and $n$ where all eigenvalues are real.
		Solving equation \eqref{eq:char} gives three eigenvalues $\lambda_1,\lambda_2,\lambda_3$ as functions of $m,n$.
		Let $\lambda_1$ be the largest of the eigenvalues.
		
		The maximal average success probability of the $3^d\rw1$ QRAC is given by the expression
		\begeq
			\label{eq:sumlambda}
			\max\sum_{i,j,k}\lambda_1(m_{ijk},n_{ijk}).
		\endeq
		From this formula one can easily see that this depends not only on unbiasedness of the basis, $\{m_{ijk}\}$, but also on the additional parameters $\{n_{ijk}\}$.
				
		Now, we show that in order to maximize the success probability of a $3^d\rw1$ QRAC in general it is not sufficient to take arbitrary MUBs as measurements. In a $3^d\rw 1$ QRAC protocol, let us use 3 out of the $d+1$ mutually unbiased bases. There are ${d+1}\choose{3}$ such subsets. From numerical results we observe that the choice of bases subset of MUBs affects the final average success probability \eqref{eq:qasp}.
		
		As an example there are 20 distinct subsets of 3 MUBs in $d=5$. By direct calculation, using 10 of those subsets we obtain an average success probability of $0.610855$, and with the other 10 subsets $0.596449$. \textit{Hence the MUB condition alone does not guarantee the highest average success probability.} We call this rather surprising behaviour, an anomaly. We observe this effect also in higher dimensions. We present our numerical observations for higher dimensions in Tab.~\ref{fig:anomalytable}.
		
		\begin{table}
			\label{table:tabmub}
			\begin{tabular}{cc|c|c|c|c|c|c|c|l}
				\cline{3-9}
				& & \multicolumn{7}{ c| }{d} \\ \cline{3-9}
				& & 5 & 7 & 8 & 9 & 11 & 13 & 16 \\ \cline{1-9}
				\multicolumn{1}{ |c  }{\multirow{3}{*}{n} } &
				\multicolumn{1}{ |c| }{3} & \cmark & \xmark & \xmark & \cmark & \xmark & \cmark & \xmark   \\ \cline{2-9}
				\multicolumn{1}{ |c  }{}                        &
				\multicolumn{1}{ |c| }{4} &\xmark & \cmark &\xmark & \cmark & \cmark & \cmark & \cmark    \\ \cline{2-9}
				\multicolumn{1}{ |c  }{}                        &
				\multicolumn{1}{ |c| }{5} & \xmark & \xmark & \xmark & \cmark &  \cmark & \cmark & \cmark   \\ \cline{1-9}
				
			\end{tabular}
			\caption{Table showing which pairs of $n,d$ values produce different average success probabilities depending on the choice of subsets of MUBs, i.e. an anomaly. The symbol \cmark means that we observe the anomaly for this particular number of inputs and dimension and \xmark $\text{ }$ means that there is no anomaly. \label{fig:anomalytable}}
		\end{table}
		
		From our numerical analysis we have observed something else which might be of interest. Consider the subset of MUBs which give the highest observed average success probability. Here, for every pure state that Alice encodes, the success probability, $p(b=x_y)$, does not depend on the basis $y$ Bob uses to measure. In contrast, for the cases where a lower success probability was observed, the successful collapse probability of the encoded state depended on the measurement basis (and was therefore not uniform). We believe that this effect is somehow related to the complexity of higher dimension $\CC^d$ spaces. In general given $n$ vectors $\{|v_i\ra\}_i$ in $\CC^d$, it is not possible to find another normalized vector $|\psi\ra$ with equal overlap $|\la v_i | \psi\ra |^2$ to all $n$ vectors.
		
		There is no obvious pattern which could suggest in which case those anomalies are present. It is surprising for example that in 8 dimensions we did not find this effect. One of the research directions in this topic is answering the question: are there other protocols which will be affected by this sensitivity for choosing different subsets of MUBs. Another question is: is there any pattern which could help us predict which cases of $n,d$ have a so-called MUB anomaly. \\
		

\section{The Unbiasedness Measure $\bar{Q}$}
\label{sec:Qbar}

To obtain \eqref{eq:Qmeasure}, we simply substitute the expressions $\bar{P}_{c,q}(2,d)$ of the optimal classical and quantum $2^d\rw 1$ Random Access Codes. Dropping the argument $\{\psi^i\}_i$, this yields:
\begeq
		\bar{Q} =\frac{1}{\binom{n}{2}} \sum_{\{a,b\}\in S^n_2} \left( \frac{1}{d(\sqrt{d}-1)} \sum_{i,j\in[d]} \left( |\la \psi^a_i | \psi^b_j \rangle | - \frac{1}{d} \right) \right).
		\label{eq:Qmeasure2}
\endeq
We rewrite \eqref{eq:dsq}, to have a side-by-side comparison.
\begeq
\bar{D}^2 =\frac{1}{\binom{n}{2}} \sum_{\{a,b\}\in S^n_2} \left( 1 - \frac{1}{d-1} \sum_{i,j\in [d]} \left(|\la \psi^1_i |\psi^2_j\ra|^2 -\frac{1}{d}\right)^2 \right).
\endeq
Notice that $\bar{D}^2$ is a function of the ''probabilities" $|\la \psi^a_i | \psi^b_j\ra|^2$, and tries to measure how much these values differ from the uniform distribution. This has a very aesthetic mathematical interpretation. However, it lacks a clear operational meaning which our proposed measure $\bar{Q}$ posseses. Finally, we point out that $\bar{D}^2$ is not a function of $\bar{Q}$ or vice-versa, except for the trivial cases ($0$ and $1$).

\section{See-Saw for $(n,2)^6 \rw 1$ pQRACs}
\label{sec:seesaw}

Here we use the see-saw method and our unbiasedness measure ($\bar{Q}$ or $\bar{p}$) to add to the pre-existing evidence that $4$ mutually unbiased bases do not exist in dimension 6.

In the see-saw method one interlaces two steps of SDP optimizations: For every other iteration one optimizes Alice's preparation states for a given set of Bob's measurements, and for the remaining steps one optimizes Bob's measurements for a given set of Alice's preparation states \cite{seesaw1}. Even though this method does not guarantee the convergence to the global maximum, it has proved to be efficient in Bell-type scenarios, see e.g. \cite{seesaw2}.
	
	The efficiency of the see-saw method can be improved in the following way. Let $\{ M_y^b\}$ be the set of measurements of Bob. Note that when Alice gets as input the information that Bob has either $y_1$ or $y_2$ setting, her strategy should be to send a state $\rho$ maximizing the value of $\Tr \left[ \rho \left( M_{y_1}^{x_{y_1}} + M_{y_2}^{x_{y_2}} \right) \right]$. This is obtained if the state is in the subspace of vectors with maximal eigenvalue of the operator $M_{y_1}^{x_{y_1}} + M_{y_2}^{x_{y_2}}$. Thus the see-saw step optimizing the states can be replaced with a construction of the states basing on eigenvectors decomposition of the measurement, cf.~remark below \eqref{eq:qasp}.

	We have used the see-saw optimization for several cases of $(n,k)^d \rightarrow 1$ pQRACs. The results are given in Tab.~\ref{tab:seesaw}.
	
	\begin{table}[t]				
		\begin{tabular}{|c|c|c|}
			\hline 
			case $(n,m)^d$ & $\bar{D}^2$  & $\bar{Q}$ \\  \hline 
			$(4,2)^3$ & $\mathbf{0.999999}59$ &  $\textbf{0.999999}72$
 \\  \hline  
			$(5,2)^4$ & $\mathbf{0.999999}22$ &  $\textbf{0.999999}39$
 \\  \hline  
			$(6,2)^5$ & $\mathbf{0.999999}04$ &  $\textbf{0.999999}10$
 \\  \hline 
			$(3,2)^6$ & $\mathbf{0.999998}69$ &  $\textbf{0.986390}54$
 \\  \hline 
			$(4,2)^6$ & $\mathbf{0.998283}88$ &  $\textbf{0.998046}89$
\\  \hline  
			$(4,2)^7$ & $\mathbf{0.992371}97$ &  $\textbf{0.977929}16$
 \\  \hline 	
		\end{tabular}
		\caption{The results of execution of the see-saw optimization for different pQRACs in search of $n$ MUBs in dimension $d$. The table shows the value of $D^2$ parameter for measurements obtained in optimizations and the success probability of the pQRAC with these measurement basis.
		We conclude that for $d = 3,4,5$ we have found the maximal amount of MUBs using the method. \label{tab:seesaw}}
	\end{table}

	Above we observed that for the $(4,2)^6$ pQRAC game to obtain the maximal guessing probability \eqref{eq:Qmeasure}, the measurements of Bob has to be unbiased, and the value of \eqref{eq:dsq} is $1$. We will now use the Monte Carlo method to investigate the relation between the quantities \eqref{eq:dsq} and the games' average success probability \eqref{eq:Pmeasure}. Our results show that the majority of highly unbiased measurements in the meaning of \eqref{eq:dsq} gives large values of the $(4,2)^6$ pQRAC game, \eqref{eq:Pmeasure} and vice versa.
	
	We have randomized instances of measurements of Bob in the following way. For each measurement we randomize a $6 \times 6$ complex matrix $A$ with each entry given by uniform distribution on the set $\{ x + \iu y: x,y \in [0,1] \}$; then the SVD decomposition is performed giving matrices $U$, $S$ and $V$ satisfying $U S V^{\dagger} = A$, with $U$ and $V$ being unitary. We form the basis for the measurement by taking subsequent columns of the $U$ matrix.
	
	In our Monte Carlo experiment we randomized $10 000$ instances of random sets of measurements, calculated optimal states and the value of the $(4,2)^6$ pQRAC game, $\bar{p}$, \eqref{eq:Pmeasure}, and the value of $\bar{D}^2$, \eqref{eq:dsq}. 
	
	\begin{figure}
		\includegraphics[scale=0.6]{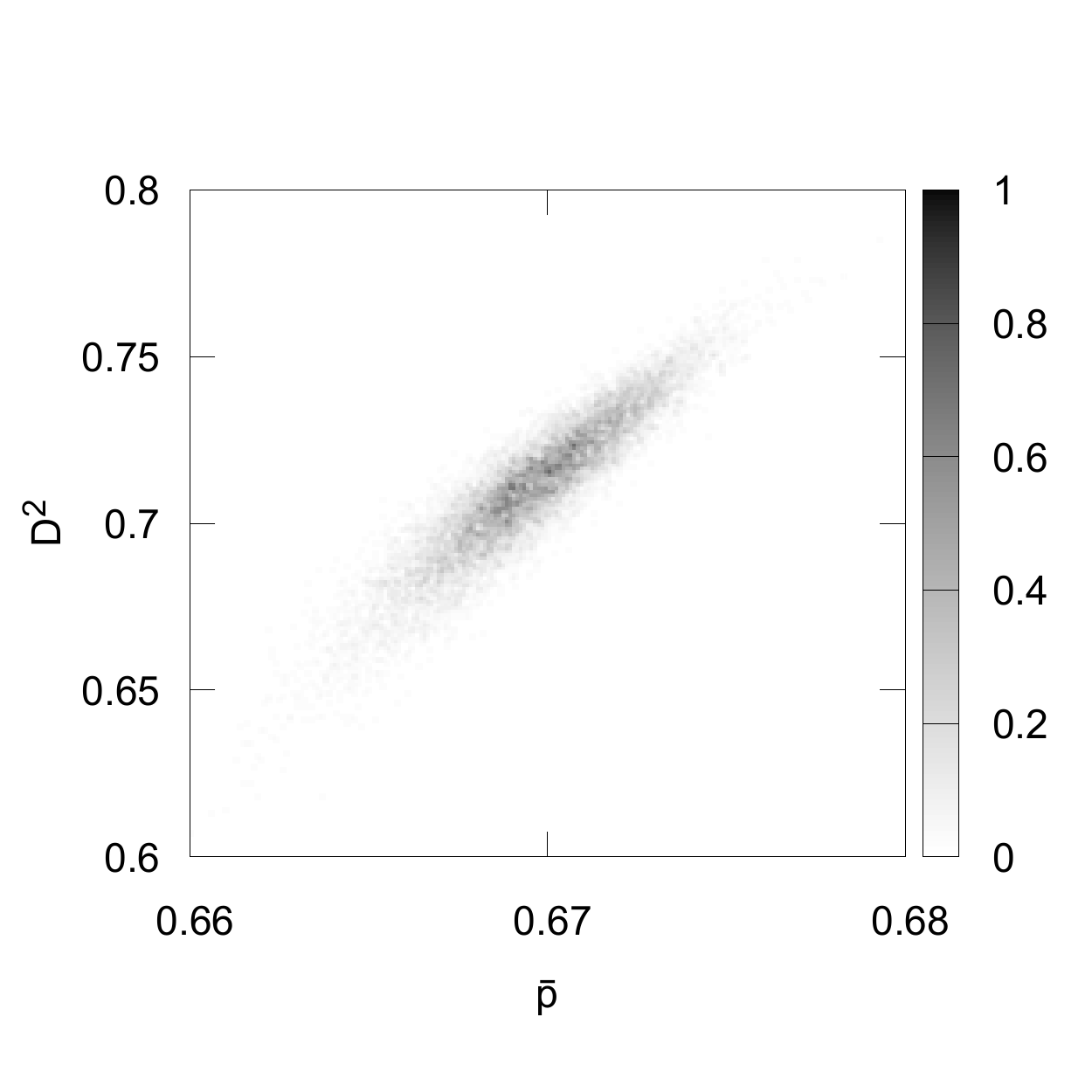}
		\caption{The values represents relative densities of the number of different basis giving similar values of the pair $(\bar{p}, D^2)$. \label{fig:HSplot}}
	\end{figure}

	The meaning of this result is the following. If one considers a set of basis constructed in a natural way, then he can expect that the higher the unbiasedness, $D^2$, the higher the success probability $\bar{p}$ in the $(4,2)^6$ pQRAC game, and vice versa. 
	
 The Spearman \cite{Spearman1,Spearman2} rank correlation coefficient is a way of measuring the rank correlation between two (random) variables. It is a generalization of the Pearson coefficient, where the latter measures only linear dependence, and the former is able to express any monotonic relationship, with $0$ meaning no correlation and $\pm1$ meaning perfect monotonicity. We have calculated the value of Spearman rank correlation coefficient for the randomized data points and obtained that it's value is about 0.91 meaning very strong dependence, but also displaying that these measures are subtly different.

\section{Estimating the Memory Requirements of the Navascues Vertesi hierarchy}
\label{sec:estimate}
The random moment matrices $\Gamma^j_k$ are created by generating $A=\binom{n}{2} d^2$ random pure states $\{\rho^j_{\mathbf{x},z}\}$, and $B=nd$ projective measurement operators $\{M^{y,j}_b\}$.  Let $w_k =  w_k(\{ \{\rho^j_{\mathbf{x},z}\},\{M^{y,j}_b\}\})$ be the vector containing all strings of operators of length smaller than or equal to $k$. For convenience we relabel the indicess of the states $\{\rho^j_\alpha\}_{\alpha =1}^A $, and of the measurements $\{M^j_\beta\}_{\beta =1}^B$. Then, e.g.\ $w_1 = (\mathbbm{1}, \rho^j_1 , \ldots , \rho^j_{A} , M^{j}_1,\ldots , M^{j}_B)$, and $w_2$ would further contain the elements ($\rho_{\alpha_1}^j \rho^j_{\alpha_2}$, $\rho^j_{\alpha_1}M^j_{\beta_1}$, $M^j_{\beta_1}\rho^j_{\alpha_1}$, $M^j_{\beta_1}M^j_{\beta_2}$), $\forall \alpha_{1,2}\in [A] , \beta_{1,2}\in [B]$. The moment matrix is constructed as follows:
\begeq
\left( \Gamma^j_k \right)_{i,j} = \tr \left[ (w_k)_i^{\dagger} (w_k)_j \right] ,
\endeq
with $\left( \cdot \right)_{i,j}$  corresponding to the components of the matrix, and similarly for vectors. We note that by construction $\left( \Gamma^j_k \right)_{i,j}  = \left( \Gamma^j_k \right)_{j,i}$, and $\Gamma^j_k \geq 0$. 

We wish to estimate the size of $\SSS^k_d$. The moment matrices are $|w_k|$-by-$|w_k|$, and by construction we see that the length of the vectors $|w_k| = \sum_{i=0}^{k} (A+B)^i = O \left( (A+B)^k \right)$. In general, it is unknown how many matrices $\Gamma^j_k$ need to be generated until the set $\{\Gamma^j_k\}_j$ spans the full space $\SSS^k_d$. We can upper bound this number by $\frac{1}{2} |w_k|^2$ trivially (since the $\Gamma^j_k$ are real-symmetric). We note that e.g.\ for a $(4,2)^5\rw 1$ pQRAC the actual number of linearly independent matrices was just 6.5\% lower than this trivial bound. Putting everything together, we see that the space $\SSS^k_d$ is described by roughly $O \left( (A+B)^{4k}\right)$ real parameters. 

For the case of 4 MUBs in dimension 6, $(A+B)=240\approx 2^8$. The $k$-th level of the hierarchy thus requires roughly $2^{32k}$ bits of memory just for indices and a further $2^6$ bits if each number is stored in double precision floating-point format. This exponential growth on memory requirements means that a typical 32-bit operating system (or mathematical software) cannot compute the first level of the hierarchy since it will run out of available memory indices. In fact, we encountered this problem when executing the program without symmetries in 32-bit OCTAVE. More drastically, a 64-bit system would be unable to calculate the second level, and all of the world's storage space is insufficient for the third level (approximately $2^{74}$ bits of information were generated in 2016 \cite{worlddata}).

\textit{Requirements considering symmetries.-} Even if we restrict our optimization to the symmetric subspace $\mathcal{G}(\SSS^k_d)$, the moment matrices are still of size $|w_k|$-by-$|w_k|$. Just as we do not know a priori the dimension of the space of feasible moment matrices, $\text{dim}(\SSS^k_d)$, we cannot predict the dimension of the symmetric subspace, $\text{dim}(\mathcal{G}(\SSS^k_d))$. Therefore, our proposal can at most be a quadratic improvement, i.e.\ of $O \left( (A+B)^{2k}\right)$. However, as we have portrayed, this reduction is sufficient to calculate the first level of the hierarchy as well as a subset of the $Q^{1+AB}$ level.

\section{Symmetries}
\label{sec:Symmetries}

We find it easier to describe the symmetry group $\mathcal{G}$ by analyzing that there are two different types of possible relabelings. Let Bob have $n$ different work stations, and on each station he has a quantum measurement device. The success probability should not depend on which physical device is situated at which work station, so long as Alice is aware of which measurement basis is being used for every input $y$ of Bob. Likewise, Bob is free to relabel the outputs of every work station at will without dropping the success rate if Alice applies the same permutation to her inputs. 

In order to properly describe the group, we will use the picture where Bob first moves the measurement apparatuses around with a permutation $\omega \in \mathcal{S}_n$, and \textit{afterwards} he relabels the device in work station $i$ with the permutation $\pi_i \in \mathcal{S}_d$. We stress that the permutations $\pi_1,\pi_2,\ldots$ address a specific work station - as opposed to a specific measurement device. Then, any relabeling of device outputs, and devices themselves can be achieved with these operations. We symbolically denote this abstract group element $g\in \mathcal{G}$ as:
\begin{equation}
g = \pi_1 , \pi_2 , \ldots , \pi_n ; \omega  .
\end{equation}
If, $\tilde{g} = \tilde{\pi}_1,\ldots,\tilde{\pi_n};\tilde{\omega} \in \mathcal{G}$, then $\tilde{g} g \in \mathcal{G}$:
\begin{equation}
\tilde{g}g = \tilde{\pi}_1\pi_{\tilde{\omega}^{-1}(1)} , \ldots , \tilde{\pi}_n\pi_{\tilde{\omega}^{-1}(n)} ; \tilde{\omega}\omega . 
\end{equation}
The inverse of $g$ is:
\begin{equation}
g^{-1} = \pi^{-1}_{\omega^{-1}(1)}, \pi^{-1}_{\omega^{-1}(2)} , \ldots , \pi^{-1}_{\omega^{-1}(n)} ; \omega^{-1}.
\end{equation}
More importantly though, is to see the action of $g$ on Bob's measurements and Alice's encoded states:
\begin{subequations}
\begin{align}
g(M^y_b) &\mapsto M^{y'}_{b'} \\
g(\rho_{x_{z_1},x_{z_2},\{z_1,z_2\}}) &\mapsto \rho_{x'_{z_1'},x'_{z_2'},\{z_1',z_2'\}}) ,
\end{align}
\end{subequations}
such that
\begin{align*}
b' &= \pi_{y'} (b) \\
y' &= \omega(y) \\
x'_{z_i'} &= \pi_{z_i'} (x_{z_i'}) &\SP \forall i=1,2 \\
z_i' &= \omega(z_i) &\SP \forall i=1,2.
\end{align*}
It is therefore clear, that if we apply an element $g$ to the vector of operator strings $w_k$, it just acts as a permutation. The operators are the same, just re-indexed. Hence, we can represent the elements $g$ as the permutation matrices $G$. What is more, by construction we know that there is an order in which these elements are applied, and $G$ can be seen as a product of permutation matrices:
\begin{equation}
G = \Pi_1 \Pi_2 \cdots \Pi_n \Omega ,
\end{equation}
where $\Pi_i$ is the matrix representation of $\pi_i$, and similarly $\Omega$ is the matrix representation of $\omega$. When this representation acts on a moment matrix, it will first shift the indices of the measurement devices with $\Omega \Gamma^j_k \Omega^T$, and later it will reshuffle the output labels of each work station. Notice that the representations of the output labels commute amongst each other ($[\Pi_i,\Pi_j]=0$), but not with $\Omega$. We are now in a position to calculate the average of the moment matrix over the group representation \eqref{eq:avgsymm}.
\begin{equation}
\label{eq:avgsymmDecomp}
\hat{\Gamma}^j_k = \frac{1}{|\mathcal{S}_n||\mathcal{S}_d|^n } \sum_{\Pi_1}\cdots \sum_{\Pi_n} \sum_{\Omega} \Pi_1 \cdots \Pi_n \Omega \Gamma^j_k \Omega^T \Pi_n^T \cdots \Pi_1^T
\end{equation}
Crucially for implementations, the most important thing about \eqref{eq:avgsymmDecomp} is that the sum may be rewritten as:
\begin{equation}
\sum_{\Pi_1} \Pi_1 \left( \cdots \left(\sum_{\Pi_n} \Pi_n \left( \sum_{\Omega} \Omega \Gamma^j_k \Omega^T \right)\Pi_n^T  \right) \cdots \right) \Pi_1^T .
\end{equation}
This reduces the amount of operations needed to perform the average from $|\mathcal{S}_n||\mathcal{S}_d|^n$ to $|\mathcal{S}_n|+n|\mathcal{S}_d|$. These methods apply to a wide variety of SDP relaxation problems, and will be further discussed in \cite{symEP} in a more rigorous fashion.


	\end{appendix}
	
\end{document}